\documentclass[10pt,sigconf]{acmart}

\usepackage{booktabs} 

\graphicspath{{figure/}{figures/}}

\copyrightyear{2021}
\acmYear{2021}
\setcopyright{acmlicensed}\acmConference[IWCI'21]{Interdisciplinary Workshop on (de) Centralization in the Internet}{December 7, 2021}{Virtual Event, Germany}
\acmBooktitle{Interdisciplinary Workshop on (de) Centralization in the Internet (IWCI'21), December 7, 2021, Virtual Event, Germany}
\acmPrice{15.00}
\acmDOI{10.1145/3488663.3493691}
\acmISBN{978-1-4503-9138-2/21/12}

\begin{document}
\title{Centralization is about Control, not Protocols}
\subtitle{Position Paper}

\author{Henning Schulzrinne}
\orcid{0000-0001-7604-1122}
\affiliation{%
  \institution{Columbia University}
  \streetaddress{1214 Amsterdam Ave}
  \city{New York} 
  \state{NY} 
  \postcode{10027}
}
\email{hgs@cs.columbia.edu}

\renewcommand{\shortauthors}{H. Schulzrinne}

\begin{abstract}
Many common ``consumer'' applications, i.e., applications widely used by non-technical users, are now provided by a very small number of companies, even if that set of companies differ across geographic regions, or rely on a very small number of implementations even if the applications are largely standards-based. While likely only a partial solution, we can draw on earlier regulatory experiences to facilitate competition or at least lessen the impact of the lack thereof.
\end{abstract}

%
%




\maketitle

\section{Introduction}

During the early years of internet, the evolving network and applications could be seen as an antidote to traditional communication systems and media: Instead of AT\&T, thousands of internet service providers offered access. Instead of hearing only the opinions of syndicated columnists or news anchors, tens of thousands of newsgroups provided everyone with a voice. Any email server could send messages to any other server and expect it to be delivered to the recipient. As with other electronic communication systems before (see T. Wu, \emph{The Master Switch} \cite{Wu10:Master}), including radio and television, this state of decentralized operation and content dissemination did not last. In most countries, key layers of the stack generally have, at most, three providers that command most of the market, whether offering the choice between DSL, fiber and cable for access, two mobile operating systems, three mobile providers, or AWS, Google Cloud or Azure for cloud service at the infrastructure level or consumer email services, search engines, feed-based social media companies, or video content distributors for applications and content. By itself, this is not surprising -- many mature industries, whether airlines, beer \cite{Hils2107:Biden} or hospitals, have experienced decreases in competitive choices \cite{Grul1810:US}, but the internet was supposed to be different.

Here, I want to emphasize that what matters is centralization of \emph{control}, not protocol design. To take an old example, if AWS were to run S3 as a DHT, that may yield advantages of resiliency, but does not decrease centralization of cloud-based storage services, i.e., the ability of users to materially affect the design and operation of that service or to choose the lowest-price or highest-performance option among many competing offerings.

When considering options for alternatives, advocates should be clear on whether they envision moving from a single dominant provider to, say, three providers, or whether they would only consider their goal met once every household runs their own services and every community runs their own network. Often, the latter is technically more interesting, but may have such high efficiency penalties that such systems are difficult to sustain economically. Also, it helps to clearly identify the cost of concentration such as higher consumer prices or lack of privacy options. 

\section{How Do We Measure Centralization and Concentration?}

In economics, the standard metric for market concentration is the Herfindahl-Hirschman Index (HHI), which sums up the squares of percentage market shares. It can range from near zero to 10,000, with that value indicating that one company has a 100\% market share. The US Department of Justice considers post-merger HHI values of 1,800 and above as ``highly concentrated'' during merger reviews \cite{DOJ9704:Horizontal}. For example, for desktop search, the HHI is roughly 7,800, with Google having a market share of 87\%. Estimating market share can depend strongly on defining the market, which is often the most contentious part of the discussion: Are Google and Amazon part of the same search market?

\section{Users Prefer Services over Systems}

The classical model of (IETF) protocol design assumed a version of the waterfall model: Standards organizations provide standards that are sufficiently detailed to allow a multitude of competing open-source and closed-source implementations to interoperate. These implementations are then acquired by entities, from consumers to carriers, that might (modestly) customize and integrate these solutions into their personal or organizational workflows and service offerings. This model obviously directly derived from the earlier telecommunications industry structure. In most cases, these users, whether individuals or enterprises, do not care about the systems themselves, and have neither the skills nor interest to install, host or operate these services. Thus, they rely on operators, with the usual economies of scale and scope, leading to the operational centralization mentioned earlier. Lowering complexity and creating modular systems may make it possible for smaller, local operators to offer services. As an example, fiber networks in rural areas, operated by local electric co-ops, became possible because fiber optic networks, voice services and network management became commodities, which co-ops could purchase from larger providers. On the other hand, any model that assumes that most individuals or enterprises will want to run their own servers or otherwise care for and feed internet services is likely doomed to niche status and, in the worst case, blaming the victim. (We find such examples in exhortations for delegating the responsibility for security and privacy to end users.)

\section{A Regulated Monopoly is (Usually) Better than an Unregulated One}

Traditionally, early internet advocates had strong libertarian tendencies. Often, the fear of any regulatory intervention, assumed inherently flawed, then led to the concentration and lack of any recourse against exploitative behavior by the monopolists. Thus, instead of simply discussing how nice it would be if competition could solve problems, one might apply standard regulatory remedies to key components of the internet stack. Fortunately, we have evidence that a set of approaches seem to have worked reasonably well for these older networks, including identifier portability, protection of user data, data portability, mandatory interconnection, transparency, or prohibition against favoring one's own services. We discuss only a few here as examples; they illustrate that problems of concentration are not new and that regulatory remedies can at least blunt their impact on consumers. In the European regulatory tradition, many of these remedies are only applied to markets or market participants with significant market power.

Identifier portability allows a consumer to move their communication service to another provider, without losing the ability to be reached by others. For US telephone networks, the ability to port numbers \cite{47CFR-C} between landline and mobile providers, including across modalities, enabled competition, particularly by mobile and VoIP as new entrants. Conversely, tying the email address to a provider has made such portability difficult to achieve for consumer internet email services. But mandatory forwarding for dominant providers along with new, provider-independent naming mechanisms,  possibly scoped to a country, may help. 

A right to interconnection is enshrined for traditional voice services in the Communications Act \cite{47USC251}, informed by the early-20th century monopolistic behavior of the Bell System. Similar obligations could be applied to messaging applications, for example, and possibly the interconnection of social networks.

Among the earliest privacy regulations for electronic data were the protections for customer proprietary network information (CPNI) in the telephone network \cite{47USC222}, strictly limiting how the carrier could use the data it naturally acquired in the course of providing services to their customers.

\section{For Crucial Components, Stakeholders, not just Shareholders, Should Have Influence}

Discussions often assume that only shareholders, or, in many cases, supervoting founders, have influence over how systems are operated. However, many key infrastructures, whether access networks or social media services, could be operated as cooperatives, i.e., with users having voting rights. In other cases, corporate governance models in some European countries may offer alternatives where the voices of stakeholders other than shareholders can guide key decisions that affect these users. It is likely that such stakeholder governance would require a government mandate. The Facebook Oversight Board is an experiment in that direction, albeit with no formal or legal authority.

\bibliographystyle{acm}
\bibliography{iwci} 

\end{document}